\documentclass[11pt, a4paper]{article}
\usepackage{moriond,epsfig}
\newcommand{\HI}{\mbox{H\hspace{0.2em}{\scriptsize I}}}
\newcommand{\HII}{\mbox{H\hspace{0.2em}{\scriptsize II}}}

\newcommand{\al}{\mbox{$^{26}$\hspace{-0.2em}Al}}

\newcommand{\Msol}{\mbox{$M_{\sun}$}}

\newcommand{\pcmq}{\mbox{cm$^{-2}$}}

\newcommand{\psec}{\mbox{s$^{-1}$}}

\newcommand{\um}{\mbox{$\mu$m}}
\newcommand{\funit}{\mbox{ph \pcmq \psec}}

\newcommand{\sun}{\hbox{$\odot$}}

\def\la{\mathrel{\hbox{\rlap{\hbox{\lower4pt\hbox{$\sim$}}}\hbox{$<$}}}}
\def\ga{\mathrel{\hbox{\rlap{\hbox{\lower4pt\hbox{$\sim$}}}\hbox{$>$}}}}

\def\deg{\hbox{$^\circ$}}

\begin{document}
\vspace*{4cm}
\title{THE CYGNUS X REGION - A MULTI WAVELENGTH VIEW}

% Changed 040623 - All authors go here
\author{J. KN\"ODLSEDER}
\address{Centre d'\'Etude Spatiale des Rayonnements, 9, avenue Colonel-Roche,
         B.P. 4346, 31028 Toulouse Cedex 4, FRANCE}

\maketitle\abstracts{ 
The Cygnus~X region is one of the most nearby massive star forming 
regions within our Galaxy, recognised by prominent emission 
throughout the entire electromagnetic spectrum, from radio to gamma-ray 
waves.
This paper reviews our current knowledge about this region by describing
its multi-wavelength characteristics.
Particular emphasis will be given to the central stellar cluster 
Cyg~OB2 that dominates the energetics and kinematics in the area.
The cluster is also believed to be an active site of nucleosynthesis, 
as traced by the observation of the 1809 keV gamma-ray line, attributed 
to radioactive decay of \al.
New observations obtained by the SPI telescope onboard the INTEGRAL 
gamma-ray observatory will be presented that corroborate this 
hypothesis, and that for the first time allow to measure the 
kinematics of the gas in the hot Cygnus~X superbubble.
}
\noindent
{\small {\it Keywords}: \HII\ regions -- ISM: bubbles -- Open clusters 
and associations: Cyg~OB2 -- nucleosynthesis}

%%%%%%%%%%%%%%%%%%%%%%%%%%%%%%%%%%%%%%%%%%%%%%%%%%%%%%%%%%%%%%%%%%%%%%%%%%%%%%%%
% Introduction
%%%%%%%%%%%%%%%%%%%%%%%%%%%%%%%%%%%%%%%%%%%%%%%%%%%%%%%%%%%%%%%%%%%%%%%%%%%%%%%%
\section{Introduction}

Massive star forming regions are impressive building blocks of active
galaxies.
They provide large amounts of UV photons leading to the ionisation of 
the interstellar medium (ISM), observable in the radio and microwave 
domains, and visible through H$\alpha$ recombination line emission.
They are important sources of interstellar dust heating, leading to 
ubiquitous infrared emission.
The strong winds of their massive stars and the subsequent supernova 
explosions release considerable amounts of kinetic energy, creating 
a rarified hot ($\sim10^6$ K) superbubble emitting in the X-ray domain;
its cavity may eventually be discerned from \HI\ and CO observations 
of interstellar gas.
Massive star forming regions are potential sources of cosmic-ray particle 
acceleration, observable by radio synchrotron emission and high-energy
gamma-ray emission.
Their massive stars synthesize considerable amounts of fresh heavy 
nuclei that are released either by the stellar winds or the 
subsequent supernova explosions, and that contribute to the chemical 
enrichment of the host galaxy.
Nucleosynthesis products may be traced by co-produced radioactive 
isotopes that are observable through their characteristic gamma-ray 
line emission.

Studying massive star forming regions in a single waveband 
provides certainly interesting clues on their characteristics; yet
a comprehensive understanding of the phenomenon requires a rigorous 
multi-wavelength approach.
I present in this paper a multi-wavelength view of the Cygnus~X region, 
one of the most nearby galactic massive star forming regions, at a 
distance of $\sim1.4$ kpc.
Cygnus~X lies in the galactic plane at longitudes $\sim80\deg$, making 
it a fairly well isolated source on the sky.
Yet, foreground dust obscuration produces substantial extinction of 
parts of the region, preventing thus a comprehensive survey in the 
visible and soft X-ray bands.
Eventually, a spiral arm structure seen tangentially over 
distances $\sim1-4$ kpc may also be present in this area of the sky, yet 
highly uncertain distance estimates make it difficult to assess the 
reality of this feature.
In any case, most massive stars in Cygnus~X, and in particular the young 
massive cluster Cyg~OB2, are situated at about the same distance 
($\sim1.4$ kpc), suggesting a physical relation of the objects in the 
region.

The proximity of Cygnus~X and its isolation from the galactic ridge 
makes it a brilliant test case for understanding massive star forming 
regions.
Cygnus~X is like a {\em Rosetta stone} of massive star formation, enabling 
a profound understanding of the various processes at work and of their 
interplay.

%%%%%%%%%%%%%%%%%%%%%%%%%%%%%%%%%%%%%%%%%%%%%%%%%%%%%%%%%%%%%%%%%%%%%%%%%%%%%%%%
% Cygnus OB2
%%%%%%%%%%%%%%%%%%%%%%%%%%%%%%%%%%%%%%%%%%%%%%%%%%%%%%%%%%%%%%%%%%%%%%%%%%%%%%%%
\section{Cygnus OB2}

%%% Figure: 2MASS image %%%%%%%%%%%%%%%%%%%%%%%%%%%%%%%%%%%%%%%%%%%%%%%%%%%%%%%%
\begin{figure}[!t]
  \center
  \epsfxsize=15cm \epsfclipon
  \epsfbox{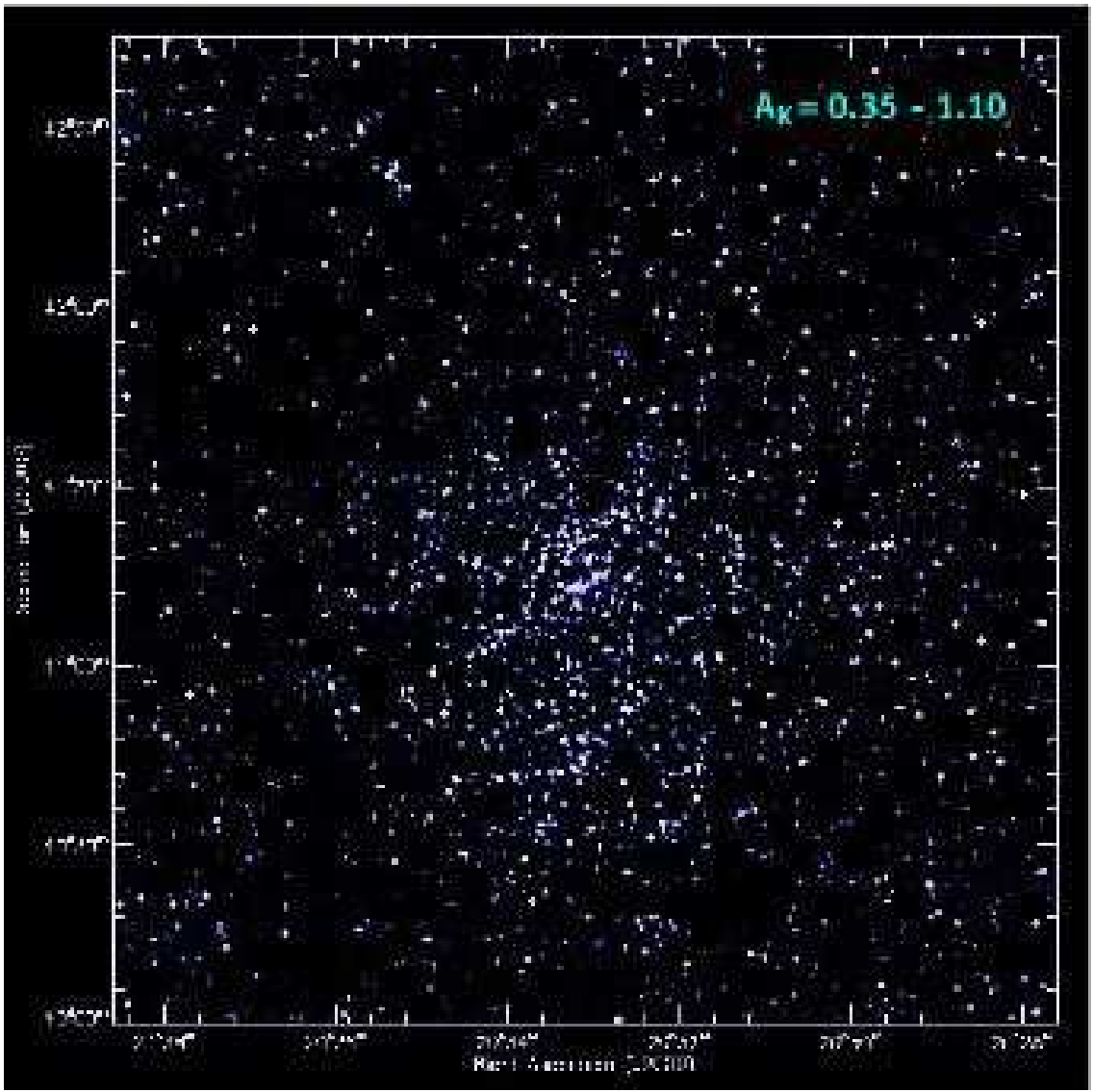}
  \caption{\label{fig:ob2-2mass}
  Synthetic absorption free plot of the stellar density distribution in 
  the Cyg~OB2 area based on 2MASS data.
  Stars in the absorption band $A_{K}=0.35-1.1$ mag are shown as symbols 
  with brightness proportional to the absolute (reddening corrected) 
  K band magnitude.
  }
\end{figure}
%%%%%%%%%%%%%%%%%%%%%%%%%%%%%%%%%%%%%%%%%%%%%%%%%%%%%%%%%%%%%%%%%%%%%%%%%%%%%%%%

Cyg~OB2, originally classified as OB association, is one of the most 
massive star clusters known in our Galaxy.
It is situated at galactic coordinates $(l,b)\sim(80\deg,1\deg)$ with an 
angular diameter of $2\deg$ (50 pc at the distance of Cyg~OB2), right at 
the heart of the Cygnus~X region.\cite{knodlseder00}
Foreground dust obscuration sheds parts of the cluster in the visible 
waveband, requiring infrared observations for a complete census.
Such a census has been obtained using the 2MASS survey, which 
revealed a total of $120\pm20$ O star members and suggests a total 
cluster mass of $(4-10) \times 10^{4}$ \Msol.\cite{knodlseder00}
A synthetic plot of the stellar distribution in the area is shown in 
Fig.~\ref{fig:ob2-2mass}.

Near-infrared spectroscopic observations support these 
findings.\cite{comeron02,hanson03}
Distance estimates to Cyg OB2 were generally situated around 1.7 kpc, 
yet the recent revision of the O-star effective temperature scale suggest 
a smaller value of 1.4 kpc.\cite{hanson03}
The cluster age has been estimated from isochrone fitting to 
$3-4$ Myr\cite{knodlseder02}, where the age range may reflect a 
non-coeval star forming event.
The total Lyman continuum luminosity of the stars has been estimated 
to $10^{51}$ ph s$^{-1}$, their mechanical luminosity amounts to
a few $10^{39}$ erg s$^{-1}$.\cite{knodlseder02,lozinskaya02}
Obviously, with such high luminosities, Cyg~OB2 should leave a clear 
imprint on the interstellar surroundings.

Cyg~OB2 houses some of the hottest and most luminous stars known in 
our Galaxy.
Among the members is a O3~If$^{\star}$ star, three Wolf-Rayet 
stars, and two LBV candidates.
Since spectroscopic measurements are still incomplete for the 
cluster, these numbers present probably lower limits to the 
population of extremely massive and evolved objects.

%%%%%%%%%%%%%%%%%%%%%%%%%%%%%%%%%%%%%%%%%%%%%%%%%%%%%%%%%%%%%%%%%%%%%%%%%%%%%%%%
% The Cygnus X region
%%%%%%%%%%%%%%%%%%%%%%%%%%%%%%%%%%%%%%%%%%%%%%%%%%%%%%%%%%%%%%%%%%%%%%%%%%%%%%%%
\section{The Cygnus X region}

%%%%%%%%%%%%%%%%%%%%%%%%%%%%%%%%%%%%%%%%%%%%%%%%%%%%%%%%%%%%%%%%%%%%%%%%%%%%%%%%
\subsection{Radio emission}

%%% Figure: Radio image %%%%%%%%%%%%%%%%%%%%%%%%%%%%%%%%%%%%%%%%%%%%%%%%%%%%%%%%
\begin{figure}[!t]
  \center
  \epsfxsize=15cm \epsfclipon
  \epsfbox{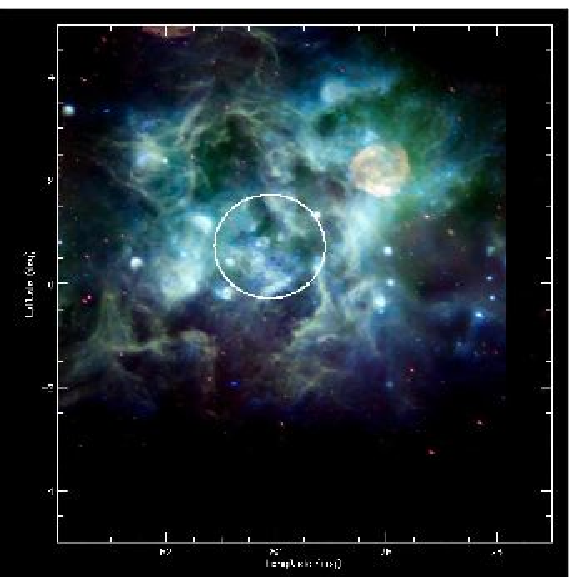}
  \caption{\label{fig:radio}
  DRAO 74 cm (red) and 21 cm (blue) composite of radio emission from 
  the Cygnus X region.
  Thermal emission appears blueish while non-thermal (i.e.~synchrotron 
  emission) appears reddish in this representation.
  The Cyg~OB2 association, represented by the white circle, is located 
  near the centre of the Cygnus~X region.}
\end{figure}
%%%%%%%%%%%%%%%%%%%%%%%%%%%%%%%%%%%%%%%%%%%%%%%%%%%%%%%%%%%%%%%%%%%%%%%%%%%%%%%%

The Cygnus~X region has been named by Piddington \& Minnett (1952)
who were the first to observe an extended source of radio 
emission in the constellation of Cygnus.
Cygnus~X is composed of numerous individual \HII\ regions\cite{downes68}
with distances estimated between $1.2-2.4$ kpc.\cite{dickel69}
In addition to these point-like sources, a component of diffuse thermal 
radio emission has been identified that constitutes $\ga 50\%$ of the 
emission in Cygnus~X.\cite{wendker70}

Figure \ref{fig:radio} presents a two-colour radio image of the region.
Thermal emission appears blueish while non-thermal (i.e.~synchrotron 
emission) appears reddish in this representation.
The dominance of thermal emission is obvious, as well as the presence 
of ubiquitous diffuse emission with embedded compact and 
ultra-compact \HII\ regions.
Only little synchrotron emission is present, such as the circular  
feature near $(l,b)\sim(78\deg,2\deg)$ which corresponds to the
$\gamma$ Cygni supernova remnant.
Apparently, the radio emission originates mainly from the ionisation of the 
interstellar medium; the supernova activity seems rather low.
These facts suggest that the Cygnus~X region is very young, probably 
not older than 4 Myr.\cite{knodlseder02}

V\'eron (1965) has suggested that Cygnus~X is a single giant \HII\ 
region powered by Cyg~OB2, yet the identification of 
individual \HII\ regions with their exciting stars in Cygnus~X
indicates that there are also other ionising sources in the area.
About $50\%$ of the ionising luminosity in Cygnus~X comes from 
Cyg~OB2, the rest is provided by massive stars that are found in the 
surrounding associations and clusters.\cite{knodlseder02}
Using typical values for electron temperature and density of \HII\ 
regions ($T_{e} = 6000$ K and $n_{e} = 10$ cm$^{-3}$, respectively), 
the ionising power of Cyg~OB2 alone would lead to a Str\"omgren sphere 
of $60$ pc in radius, corresponding to an angular diameter of $5\deg$ 
at the distance of the cluster.
This diameter is in good agreement with the observed extent of the 
diffuse thermal radio component in Cygnus~X, supporting the idea that 
this emission component represents a giant \HII\ region that is powered 
by Cyg~OB2.

%%%%%%%%%%%%%%%%%%%%%%%%%%%%%%%%%%%%%%%%%%%%%%%%%%%%%%%%%%%%%%%%%%%%%%%%%%%%%%%%
\subsection{Infrared emission}

%%% Figure: IR image %%%%%%%%%%%%%%%%%%%%%%%%%%%%%%%%%%%%%%%%%%%%%%%%%%%%%%%%%%%
\begin{figure}[!t]
  \center
  \epsfxsize=15cm \epsfclipon
  \epsfbox{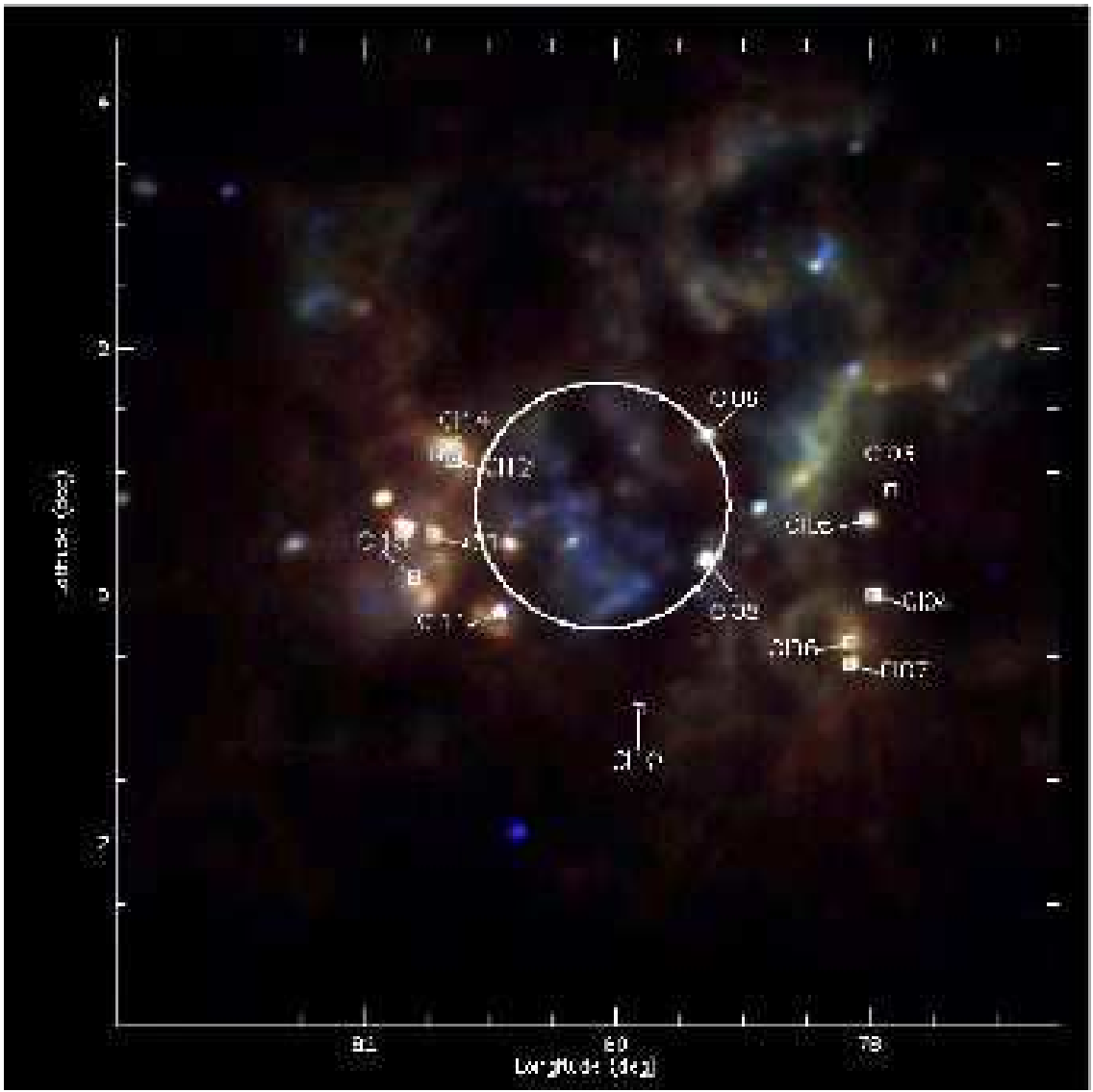}
  \caption{\label{fig:iras}
  IRAS three-colour composite of the Cygnus X region (red: 100 \um, 
  yellow: 60 \um, blue: 25 \um).
  The central region, heated by the Cyg OB2 cluster appears blue, 
  while the surrounding colder medium shows up as reddish emission.
  Bright spots coincide generally with embedded star clusters 
  (labelled according to the catalogue of embedded star clusters given by 
  Le Duigou \& Kn\"odlseder, 2002).
  }
\end{figure}
%%%%%%%%%%%%%%%%%%%%%%%%%%%%%%%%%%%%%%%%%%%%%%%%%%%%%%%%%%%%%%%%%%%%%%%%%%%%%%%%

The mid- and far-infrared image of Cygnus~X shows striking 
similarities with the radio emission: 
a diffuse emission structure with embedded point-like sources
(cf.~Fig.~\ref{fig:iras}).
The infrared spectrum becomes harder (as illustrated by the blueish 
colour) towards the centre of Cygnus~X, probably as a result of dust 
heating by the massive stars of the embedded Cyg~OB2 association.
Around Cyg~OB2, a large number of point-like sources is detected, 
that have been identified as either embedded early-type stars or star 
clusters, or young stellar objects (YSO).\cite{odenwald89}
Many of them are associated to the compact and ultra-compact \HII\ 
regions that are observed in thermal radio continuum emission.

Le~Duigou \& Kn\"odlseder (2002) have searched 2MASS 
near-infrared data for possible embedded star clusters in the area 
and found 15 such objects.
Although no detailed age information is available for these objects, 
their embedded nature suggests a fairly small age 
($\sim1$Myr).
In addition, their location at the outskirts of Cyg~OB2 indicates a 
triggered star formation event, as result of the expansion of a 
superbubble around Cyg~OB2 that has compressed the ambient ISM.

Such a superbubble has been searched for by several authors
in interstellar gas maps of the region 
(\HI\ and CO)\cite{kaftan61,heiles79,gosachinskii99,lozinskaya02},
yet velocity crowding makes the identification of interstellar bubbles 
difficult in the Cygnus area.
Assuming that the Cyg~OB2 association injects kinetic energy into 
the ISM since $\sim2$ Myr, and assuming an initial density of 100 cm$^{-3}$
(in agreement with measurements of immersed molecular clumps in 
Cyg~OB2)\cite{gredel94}, a superbubble with a radius of 63 pc should 
have been created by the association, with an actual bubble shell 
velocity of 19 km s$^{-1}$.\cite{lozinskaya02}
At the distance of Cyg~OB2, the superbubble should have an apparent 
diameter of $5\deg$, comparable to the estimated size of the Str\"omgren 
sphere, and comparable to the location of the embedded star clusters 
that may have formed in the compressed ISM surrounding the bubble.

%%%%%%%%%%%%%%%%%%%%%%%%%%%%%%%%%%%%%%%%%%%%%%%%%%%%%%%%%%%%%%%%%%%%%%%%%%%%%%%%
\subsection{The Cygnus X-ray superbubble}

The Cygnus X-ray superbubble has been discovered by Cash et al. (1980) 
as an incomplete ring of soft X-ray emission $13\deg \times 18\deg$ in 
diameter surrounding the Cygnus~X region.
The morphology of the X-ray emission is shaped by heavy 
foreground extinction due to the Great Cygnus Rift, and the underlying 
source could in reality present a nearly uniform emission 
morphology.\cite{cash80,uyaniker01}
The origin of the Cygnus X-ray superbubble is still subject to debate,
and the protagonists split into two groups who either suggest a single 
superbubble formed by Cyg~OB2\cite{cash80,abbott81} or a superposition of  
sources aligned along the local spiral 
arm.\cite{bochkarev85,uyaniker01}
Probably, the truth lies between these extreme positions: a substantial 
fraction of the X-ray emission may indeed arise from shock heating due 
to the combined stellar winds of Cyg~OB2, while other objects along 
the line of sight my also contribute to the emission.

%%%%%%%%%%%%%%%%%%%%%%%%%%%%%%%%%%%%%%%%%%%%%%%%%%%%%%%%%%%%%%%%%%%%%%%%%%%%%%%%
\subsection{Gamma-ray emission from Cygnus X}

Prominent 1809 keV gamma-ray line emission has been reported from 
the Cygnus~X region based on observations of the COMPTEL 
telescope.\cite{delrio96,pluschke01}
The 1809 keV gamma-ray line arises from the decay of \al, a radioactive 
isotope with a mean lifetime of about one million years.
\al\ is mainly produced during the core hydrogen burning phase in 
massive stars, and is subsequently ejected by stellar winds (in particular 
during the LBV and Wolf-Rayet phases) and/or supernova explosions.
The presence of \al\ in the Cygnus region is again a clear indicator 
of extensive mass loss by massive ($M>20\Msol$) stars in this area.

%%% Figure: Spectra %%%%%%%%%%%%%%%%%%%%%%%%%%%%%%%%%%%%%%%%%%%%%%%%%%%%%%%%%%%%
\begin{figure}[t]
  \center
  \epsfxsize=5cm \epsfclipon
  \epsfbox{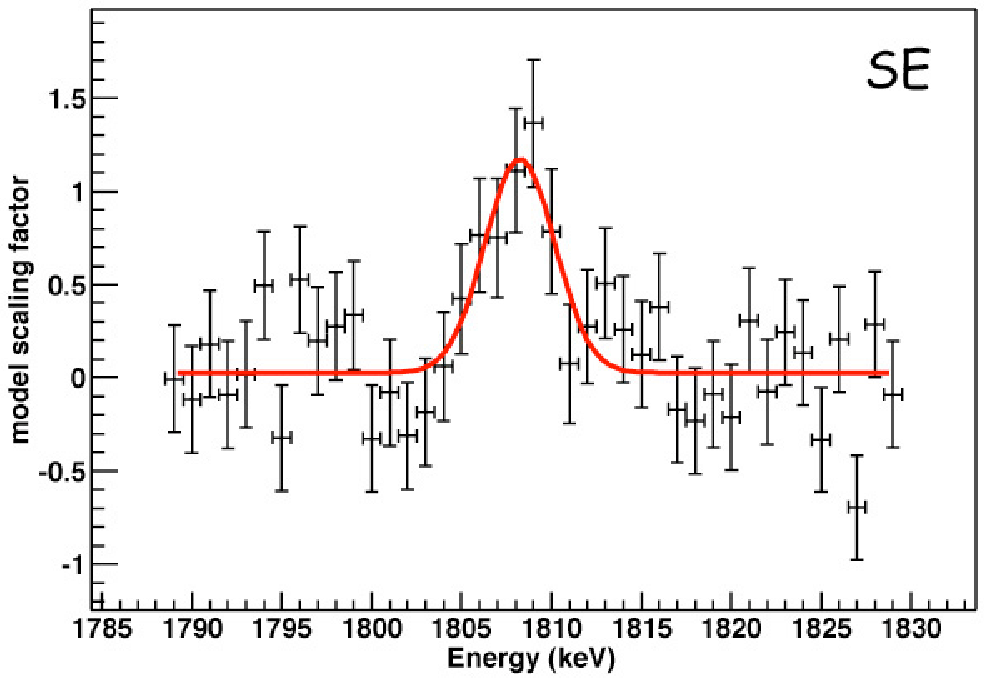}
  \hfill
  \epsfxsize=5cm \epsfclipon
  \epsfbox{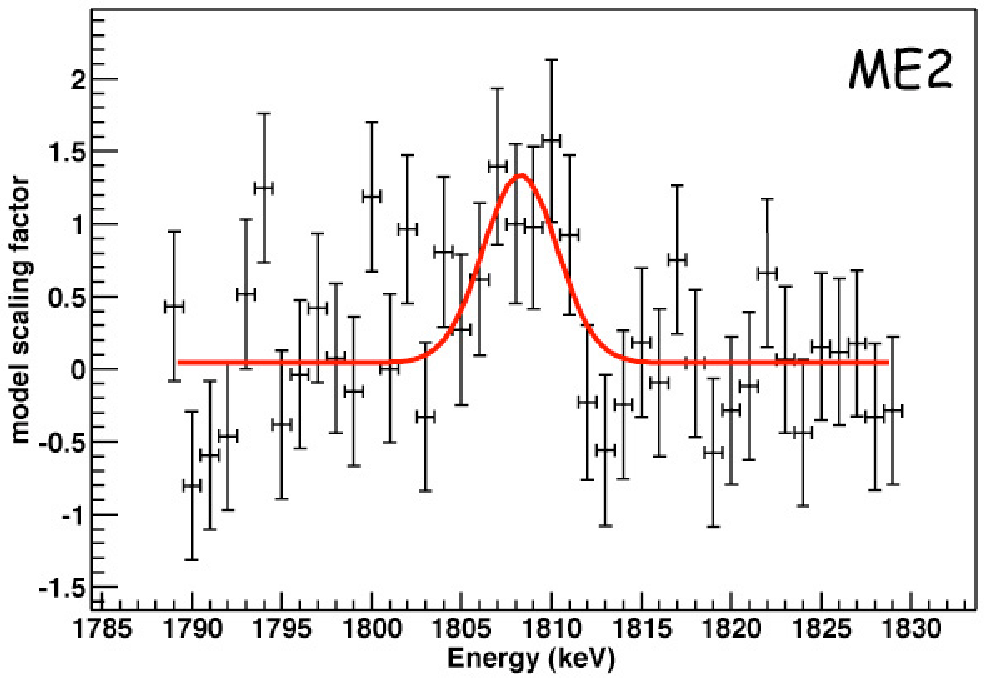}
  \hfill
  \epsfxsize=5cm \epsfclipon
  \epsfbox{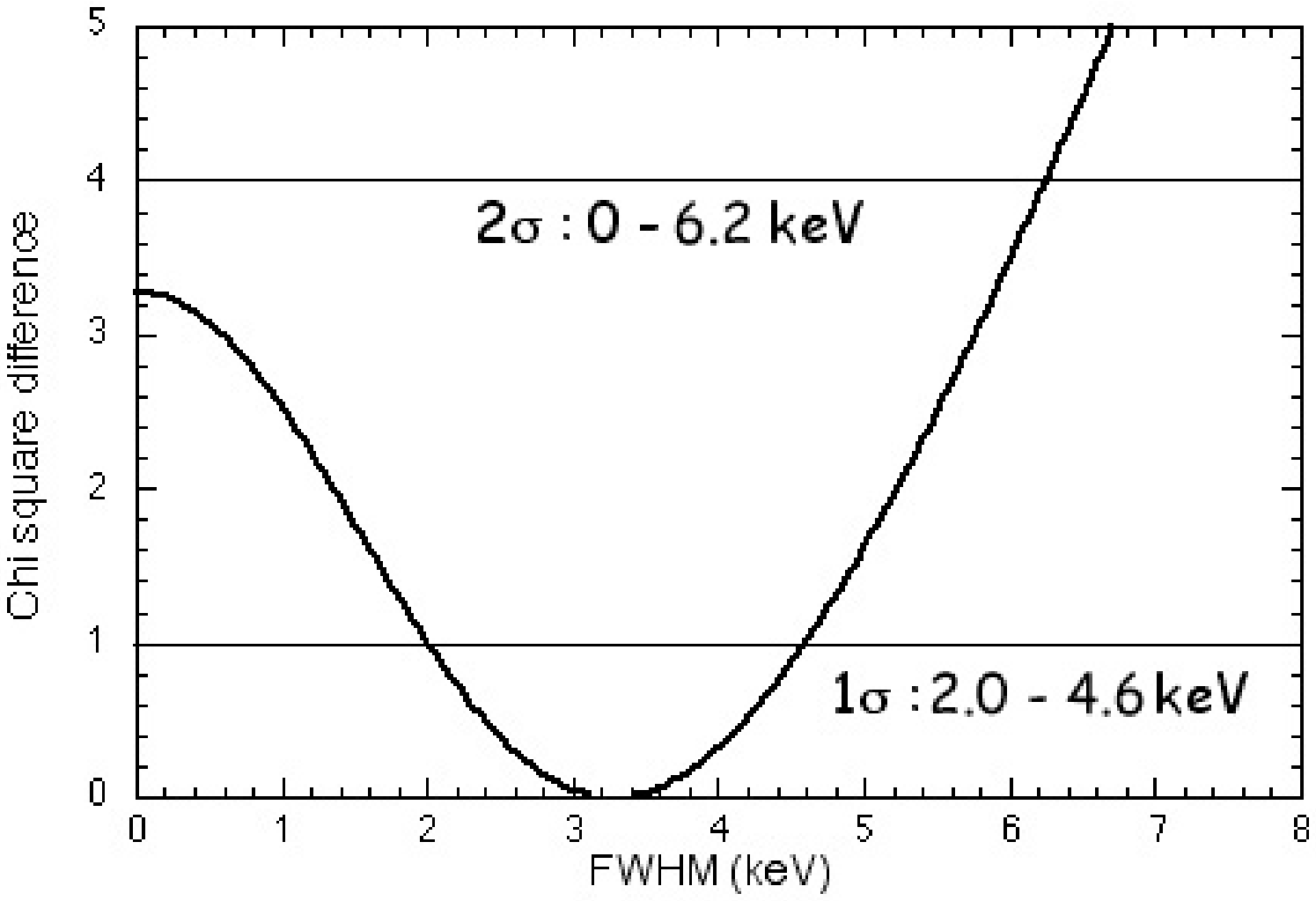}
  \caption{\label{fig:spectra}
  1809 keV gamma-ray line emission spectra obtained from observations 
  using the SPI telescope aboard the INTEGRAL gamma-ray observatory.
  The left panel shows the spectrum for single-detector events (SE), the 
  mid panel shows the spectrum for double-detector events (ME2)
  (SE and ME2 refer to two different event types 
  that are registered by the SPI telescope; the detection of the signal at 
  the same level in both event types is a valuable internal consistency 
  check of the complex data analysis; the information of both event 
  types is added to achieve the maximum sensitivity of the SPI 
  telescope).
  The right panel illustrates the $\chi^2$ statistics of a gaussian 
  shaped line profile fit as function of the astrophysical line width.
  At the $1\sigma$ level, a line broadening of $3.3\pm1.3$ keV (FWHM)
  is suggested by the data, while the line is compatible with an 
  unbroadened line at the $2\sigma$ level.
  }
\end{figure}
%%%%%%%%%%%%%%%%%%%%%%%%%%%%%%%%%%%%%%%%%%%%%%%%%%%%%%%%%%%%%%%%%%%%%%%%%%%%%%%%

The recently launched INTEGRAL gamma-ray observatory, equipped with 
the high-resolution gamma-ray spectrometer SPI, observed the Cygnus~X region 
during the performance verification phase end of 2002.
Figure \ref{fig:spectra} shows the spectra that have been obtained.
The flux integrated over the area $l=[73\deg, 93\deg]$ and $b=[-7\deg,7\deg]$
in the line amounts to $(7.2 \pm 1.8) \times 10^{-5}$ \funit.
The energy of the line is measured to $1808.4 \pm 0.3$ keV, 
compatible with the \al\ decay energy of $1808.65$ keV.
The line appears broadened, with a gaussian FWHM of $3.3\pm1.3$ keV, 
yet the statistical confidence of this measurement is still 
relatively modest.

%%% Figure: Morphology %%%%%%%%%%%%%%%%%%%%%%%%%%%%%%%%%%%%%%%%%%%%%%%%%%%%%%%%%
\begin{figure}[!t]
  \center
  \epsfxsize=5cm \epsfclipon
  \epsfbox{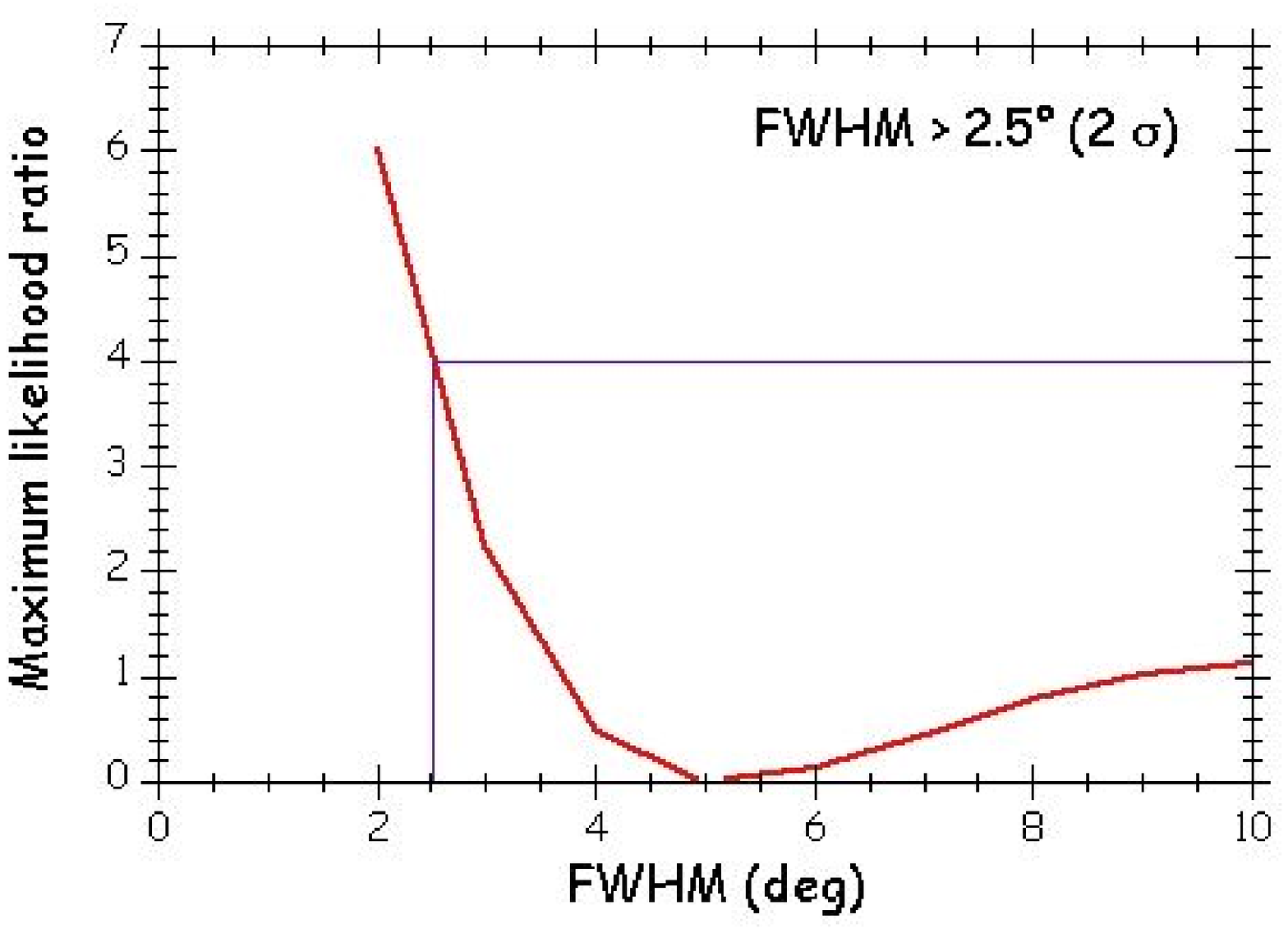}
  \hfill
  \epsfxsize=5cm \epsfclipon
  \epsfbox{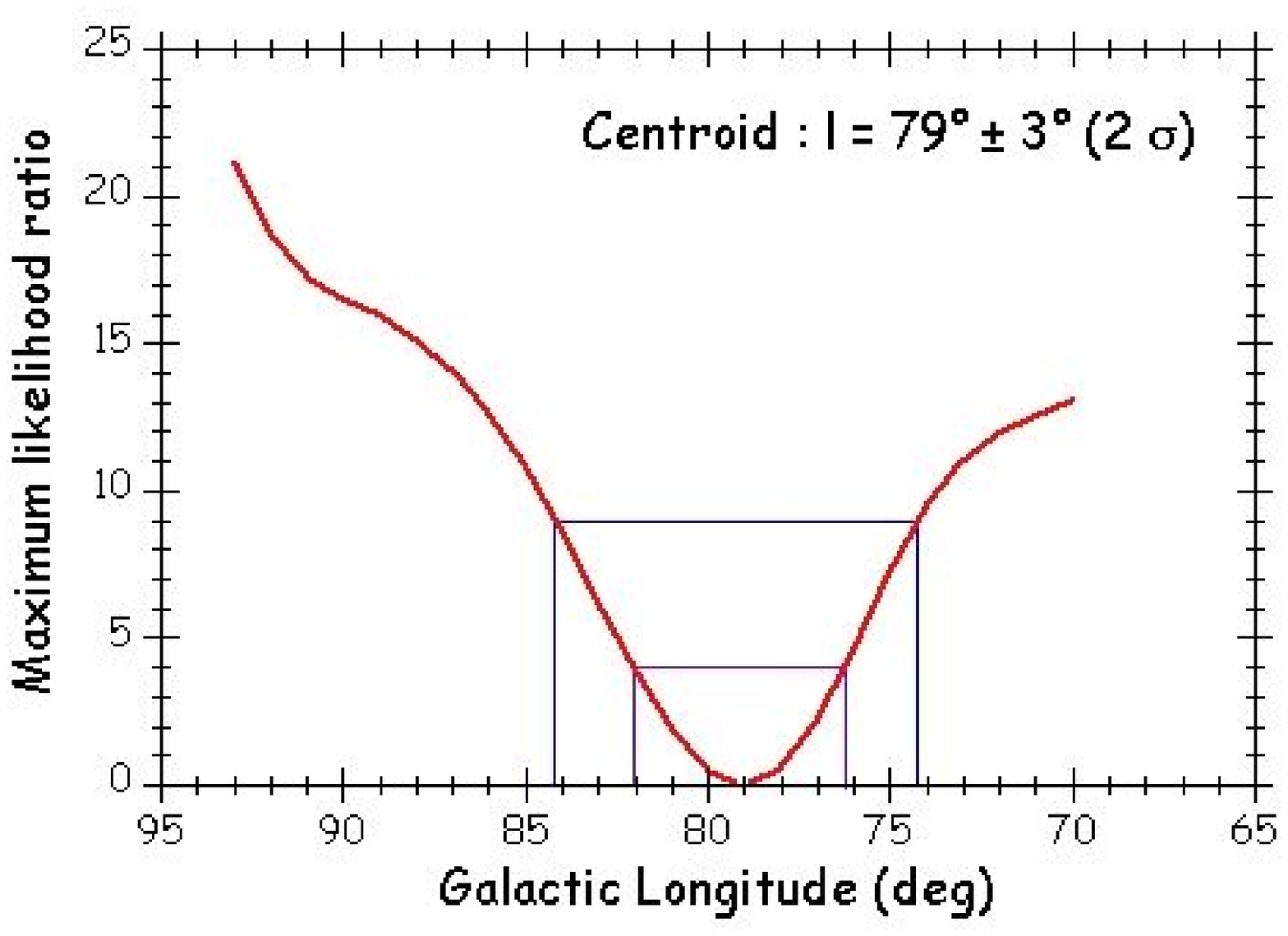}
  \hfill
  \epsfxsize=5cm \epsfclipon
  \epsfbox{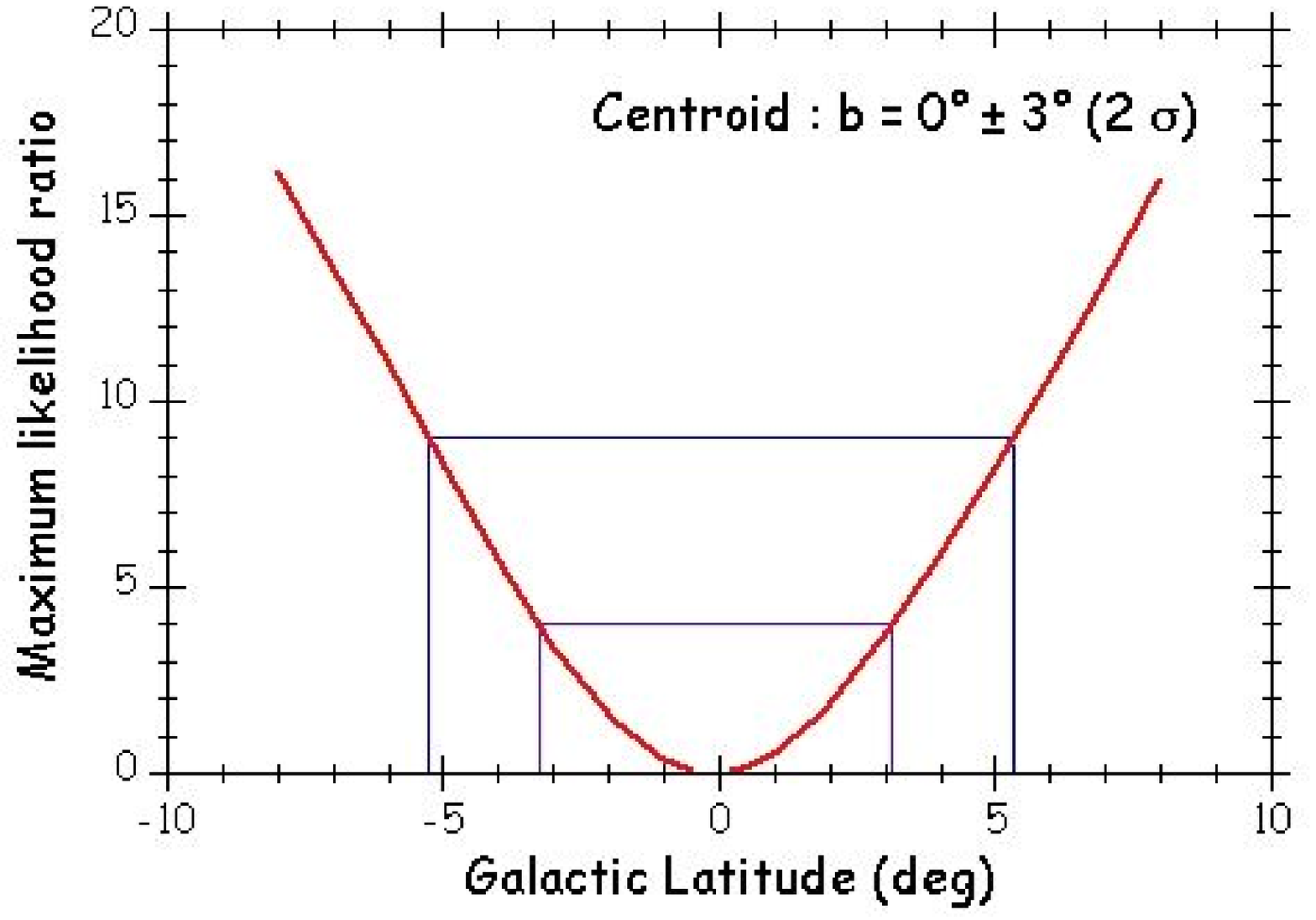}
  \caption{\label{fig:morphology}
  Morphology constraints on the distribution of 1809 keV gamma-ray line 
  emission in Cygnus~X, obtained from SPI/INTEGRAL data using the Maximum 
  Likelihood Ratio test (MLR).
  The left panel illustrates the dependence of the MLR as function on 
  the extension of the emission, assuming a 2d-gaussian shaped intensity 
  distribution centred on galactic longitude/latitude of 
  $80\deg$/$0\deg$ (the abscissa shows the FWHM of the 2d-gaussian).
  The source is extended by more than $2.5\deg$ at the $2\sigma$ 
  significance level, an optimum extension of $5\deg$ is suggested.
  The mid panel show the dependence of the MLR on
  galactic longitude, assuming a 2d-gaussian shaped intensity 
  distribution of FWHM$=5\deg$ at galactic latitudes of $0\deg$.
  The emission is located at galactic longitude $79\deg \pm 3\deg$ 
  ($2\sigma$ confidence level).
  The right panel show the dependence of the MLR on
  galactic latitude, assuming a 2d-gaussian shaped intensity 
  distribution of FWHM$=5\deg$ at galactic longitude of $79\deg$.
  The emission is located in the galactic plane at latitude $0\deg \pm 3\deg$ 
  ($2\sigma$ confidence level).
  $2\sigma$ and $3\sigma$ confidence limits are indicated as violet and 
  blue lines, respectively.
  }
\end{figure}
%%%%%%%%%%%%%%%%%%%%%%%%%%%%%%%%%%%%%%%%%%%%%%%%%%%%%%%%%%%%%%%%%%%%%%%%%%%%%%%%

SPI provided also valuable information on the morphology of the 1809 
keV emission.
Figure \ref{fig:morphology} illustrates that the emission is extended, 
with a most likely value of $5\deg$ FWHM assuming a gaussian shaped 
emission.
The emission is located at galactic longitude $79\deg \pm 3\deg$ 
and latitude $0\deg \pm 3\deg$.
Extension and location are fully compatible with the morphology of the
Cygnus~X region, and in particular, with the location of the Cyg~OB2 
cluster.

Using an evolutionary synthesis model, Kn\"odlseder et al. (2002) 
have suggested that Cyg~OB2 is indeed the dominant \al\ source in the 
Cygnus~X region.
Yet, actual nucleosynthesis models underestimate the 1809 keV line flux 
by roughly a factor of 2.
Possibly, the neglection of stellar rotation, which plays a crucial 
role for the evolution of massive stars, may explain this 
discrepancy.\cite{knodlseder02}

The SPI measurement provide for the first time an estimate of the intrinsic 
width of the 1809 keV line in Cygnus~X.
The measured width corresponds to a Doppler broadening of 
$550 \pm 210$ km s$^{-1}$ (FWHM).
This translates into expansion velocities of 170-380 km s$^{-1}$ if
a thin expanding shell is assumed, or to 240-550 km s$^{-1}$ for
a homologously expanding bubble.
Obviously, these values significantly exceed expectations (see above), 
hence it seems unlikely that the gamma-ray data trace expansion motions.
Yet, turbulent motions in hot superbubbles can reach velocities of a 
few 100 km s$^{-1}$ (De Avillez, these proceedings), and the \al\ 
ejecta may eventually follow these motions.
Hence, the gamma-ray observations may provide a direct measure of the 
turbulent motions in the hot Cygnus~X superbubble.
However, more observations of Cygnus~X are required to confirm this 
hypothesis, by providing more stringent informations on the \al\ line 
profile.

%%%%%%%%%%%%%%%%%%%%%%%%%%%%%%%%%%%%%%%%%%%%%%%%%%%%%%%%%%%%%%%%%%%%%%%%%%%%%%%%
% Conclusions
%%%%%%%%%%%%%%%%%%%%%%%%%%%%%%%%%%%%%%%%%%%%%%%%%%%%%%%%%%%%%%%%%%%%%%%%%%%%%%%%
\section{Conclusions}

Multi-wavelength observations reveal that Cygnus~X that is a textbook case 
of a massive star forming region.
Its massive (O-type) star population is catalogued using optical and 
near-infrared photometric and spectroscopic observations, unveiling a 
young massive central cluster (Cyg~OB2), surrounded by even younger 
embedded star clusters.
The youth of Cyg~OB2 explains the apparent absence of supernova 
remnants and the presence of ubiquitous diffuse thermal radio 
emission, produced by the ionising UV radiation of its O-star members.
The observations suggest that the combined stellar winds of the 
massive Cyg~OB2 members have blown a superbubble of $\sim100$ pc in diameter 
into the ISM, compressing the surrounding gas, leading to triggered star 
formation near the bubble shell.
The superbubble is filled with hot rarified gas that gives rise to soft 
X-ray emission.
The massive stars also ejected already a significant amount of 
nucleosynthesis products into the bubble, as traced by the 1809 keV 
gamma-ray line from radioactive \al\ decay.
The 1809 keV line shape indicates turbulent velocities of a few 
100 km s$^{-1}$ within the bubble that may help to accelerate cosmic 
rays in this region.
The recent discovery of a TeV gamma-ray source in Cyg~OB2 together 
with the presence of an unidentified EGRET gamma-ray source may be a 
further indication of present cosmic-ray acceleration in 
Cygnus~X.\cite{benaglia81}

%%%%%%%%%%%%%%%%%%%%%%%%%%%%%%%%%%%%%%%%%%%%%%%%%%%%%%%%%%%%%%%%%%%%%%%%%%%%%%%%
% Bibliography
%%%%%%%%%%%%%%%%%%%%%%%%%%%%%%%%%%%%%%%%%%%%%%%%%%%%%%%%%%%%%%%%%%%%%%%%%%%%%%%%

\end{document}